\documentclass[conference]{IEEEtran}
\IEEEoverridecommandlockouts

\usepackage{algorithmic}
\usepackage{graphicx}
\usepackage{textcomp}
\usepackage{xcolor}

\usepackage[utf8]{inputenc}
\usepackage[T1]{fontenc}
\usepackage{url}
\usepackage{ifthen}
\usepackage{cite}
\usepackage[cmex10]{amsmath} % Use the [cmex10] option to ensure complicance
\usepackage{amssymb}
\usepackage{amsfonts}
\usepackage{mathtools}
\usepackage{enumerate}

\usepackage{physics} % for derivatives

\usepackage{relsize} % math font size

\usepackage{tikz}
\usepackage{pgfplots}
\pgfplotsset{scaled y ticks=false}
\usetikzlibrary{arrows,shapes,backgrounds,plotmarks,positioning}
\pgfplotsset{compat=newest}                         % move axis labels close to the tick label automatically
\pgfplotsset{plot coordinates/math parser=false}
\newlength\figureheight
\newlength\figurewidth

\newtheorem{theorem}{Theorem}%[section]

\newtheorem{corollary}{Corollary}
\newtheorem{proposition}{Proposition}

\DeclareMathOperator*{\argmax}{arg\,max}

\newcommand{\Set}[1]{\{#1\}}
\newcommand{\XSet}{\mathcal{B}}
\newcommand{\X}{\mathcal{X}}
\newcommand{\Y}{\mathcal{Y}}
\newcommand{\U}{\mathcal{U}}
\newcommand{\Leak}{\mathcal{L}}
\newcommand{\PB}{\mathbf{P}}
\newcommand{\QB}{\mathbf{Q}}

\newcommand{\E}{\mathbb{E}}

\newcommand{\supp}{\text{supp}}

\def\BibTeX{{\rm B\kern-.05em{\sc i\kern-.025em b}\kern-.08em
    T\kern-.1667em\lower.7ex\hbox{E}\kern-.125emX}}
\begin{document}

\title{$\alpha$-leakage by R\'{e}nyi Divergence and Sibson Mutual Information
}

\author{\IEEEauthorblockN{Ni Ding}
%\IEEEauthorblockA{\textit{School of Computer Science}}
\textit{University of Auckland}\\
New Zealand \\
ni.ding@auckland.ac.nz
\and
\IEEEauthorblockN{Mohammad Amin Zarrabian}
%\IEEEauthorblockA{\textit{School of Computer Science}}
\textit{Australian National University} \\
 Australia \\
 mohammad.zarrabian@anu.edu.au
\and
\IEEEauthorblockN{Parastoo Sadeghi}
%\IEEEauthorblockA{\textit{School of Computer Science}}
\textit{University of New South Wales} \\
 Australia \\
 p.sadeghi@unsw.edu.au
}

\maketitle

\begin{abstract}
For $\tilde{f}(t) = \exp(\frac{\alpha-1}{\alpha}t)$,  this paper proposes a $\tilde{f}$-mean information gain measure.  R\'{e}nyi divergence is shown to be the maximum $\tilde{f}$-mean information gain incurred at each elementary event $y$ of channel output $Y$ and Sibson mutual information is the $\tilde{f}$-mean of this $Y$-elementary information gain.
Both are proposed as $\alpha$-leakage measures, indicating the most information an adversary can obtain on sensitive data. It is shown that the existing $\alpha$-leakage by Arimoto mutual information can be expressed as $\tilde{f}$-mean measures by a scaled probability.
Further, Sibson mutual information is interpreted as the maximum $\tilde{f}$-mean information gain over all estimation decisions applied to channel output.

\end{abstract}


\section{Introduction}

Information leakage was first proposed in a statistical inference framework~\cite{Calmon2012_Allerton}.
For an adversary observing published data, the information gain, or uncertainty reduction, on sensitive attribute from the prior belief (the side information obtained by the adversary) indicates the quantity of privacy leakage.
While mutual information was interpreted as the mean privacy measure in \cite{Calmon2012_Allerton,Sankar2013_TIFS}, Issa \emph{et al.} proposed maximal leakage in \cite{Issa2020_MaxL_JOURNAL} focusing on the worst-case event that incurs the maximal amount of privacy flow to the adversary.
They were later generalized by an $\alpha$-leakage~\cite{Liao2019_AlphaLeak}, tunable between average and high-risk events.
It was revealed in \cite{Ding2024_ISIT},~\cite[Theorem~1]{Liao2019_AlphaLeak} that this $\alpha$-leakage is the same as a R\'{e}nyi measure called Arimoto mutual information, where the $\alpha$-order uncertainty is quantified by R\'{e}nyi entropy~\cite{Renyi1961_Measures} and Arimoto conditional entropy~\cite{Arimoto1977} respectively for prior and posterior beliefs and the difference measures uncertainty reduction.

While the existing leakages are (essentially) using $\alpha$-order entropy measures, it is worth noting that R\'{e}nyi has also defined the $\alpha$-order relative information in~\cite{Renyi1961_Measures}, quantifying the expected uncertainty in a probability distribution in addition to another one.
It was specifically called $\alpha$-order information gain in~\cite{Sibson1969_InfRadius}, whereas the well-known name is R\'{e}nyi  divergence.
The idea is to collect the information gain, the logarithm of Radon-Nikodym derivative (also called relative information \cite[eq.(6)]{Verdu2021_ErrExp_ENTROPY}), incurred at each elementary event and get the $f$-mean of them for $f(t)=\exp((\alpha-1)t)$ w.r.t. the frequency of appearance for each elementary event.
This naturally suggests R\'{e}nyi divergence and Sibson mutual information~\cite{Sibson1969_InfRadius}, the information radius defined in terms of R\'{e}nyi  divergence, for $\alpha$-order information leakage measure.
However, existing studies~\cite{Saeidian2023_PML,Issa2020_MaxL_JOURNAL,Liao2019_AlphaLeak} only reveal that they upper bound privacy leakage of all sensitive attributes (of channel input) for fixed channel and input distribution.

In this paper, we propose R\'{e}nyi  divergence and Sibson mutual information as the exact $\alpha$-leakage of a sensitive attribute. We first define a $\tilde{f}$-mean information gain measure, where $\tilde{f}(t) = \exp(\frac{\alpha-1}{\alpha}t)$.
Viewing the posterior belief as a soft decision chosen by the adversary to estimate sensitive attribute, R\'{e}nyi divergence is shown to be the maximum $\tilde{f}$-mean information gain incurred at each elementary event $y$ of channel output $Y$. It is then proposed as $Y$-elementary $\alpha$-leakage, and the $\tilde{f}$-mean of it is measured by Sibson mutual information.
The existing leakages in \cite{Issa2020_MaxL_JOURNAL,Liao2019_AlphaLeak,Saeidian2023_PML} can be expressed by the proposed $\alpha$-leakage measures via a scaled probability distribution, by which the leakage upper bound results~\cite[Ths.1{\&}2]{Liao2019_AlphaLeak}, \cite[Th.1]{Issa2020_MaxL_JOURNAL}, \cite[Th.1]{Saeidian2023_PML} are straightforward by post-processing property.

\subsection{Notation}

Capital and lowercase letters denote random variable (r.v.) and its elementary event or instance, respectively, e.g., $x$ is an instance of $X$. Calligraphic letters denote sets, e.g., $\X$ refers to the alphabet of $X$. We assume finite countable alphabet.
Denote $P_{X}(x)$ the probability of outcome $X = x$.
For $\XSet \subseteq \X$, $\PB_{X}(\XSet) = (P_X(x): x \in \XSet)$ is a probability vector indexed by $\XSet$.
For singleton $B = \Set{x}$, $\PB_{X}(\Set{x})$ is simplified to $P_{X}(x)$.
The probability mass function $\PB_{X}(\X)$ is expressed by notation $\PB_{X}$.
The support of probability mass is denoted by $\supp(P_X) = \Set{x \colon  P_X(x) > 0}$.
Each $\PB_{X}$ is a vector in the $|\X| - 1$-dimensional probability simplex, denoted by $\triangle_{\X}$.
An optimizations over $\PB_{X}$ has the constraint set being probability simplex $\Set{\PB_X  \in \mathbb{R}_+^{|\X|} \colon\sum_{x\in\X} P_X(x) = 1 }$.
The expected value of $f(X)$ w.r.t. probability \scalebox{0.95}{$\PB_X$} is denoted by $\E_{P_X}[f(X)] = \sum_{x \in \X} P_{X}(x) f(x)$.
For the conditional probability $\PB_{Y|X} = (P_{Y|X}(y|x) \colon x\in\X, y\in\Y)$, $\PB_{Y|X=x} = (P_{Y|X}(y|x) \colon y\in\Y)$ denotes the probability of $Y$ given the outcome $X = x$.

\section{Preliminaries}

Let $f(t)=\exp((\alpha-1)t)$. The $f$-mean is $\bar{Z} = f^{-1}(\E[f(Z)])$, also called Kolmogorov-Nagumo average~\cite{Kolgomorov1930,Nagumo1930}.
Alfr{\'e}d R{\'e}nyi has defined the $\alpha$-order relative information in~\cite{Renyi1961_Measures} as an $f$-mean as follows.
For two probability distributions $\PB_X, \QB_X$ such that $\PB_X \ll \QB_X$,
the relative information for any event subset $\XSet \subseteq \X$ is\footnote{The underlying assumption for relative information in \cite[pp.553, Sec.3]{Renyi1961_Measures} is that $\QB$ refers to the original unconditional distribution of an r.v., while $\PB$ is the distribution of the same r.v. conditioned on some event. In this case, $\PB$ is absolutely continuous w.r.t. $\QB$ and the Radon–Nikodym derivative $\frac{\dd P}{\dd Q}$ is well defined.
Alfr{\'e}d R{\'e}nyi has defined a pair of entropy and relative entropy measures in~\cite{Renyi1961_Measures}: $\alpha$-order uncertainty as $f(-t) = \exp((1-\alpha)t)$-mean and $\alpha$-order $I$-relative information as $f(t) = \exp((\alpha-1)t)$-mean, later denoted by R{\'e}nyi entropy $H_{\alpha}(\cdot)$ and R{\'e}nyi divergence $D_\alpha(\cdot \| \cdot)$, respectively.
}
\begin{align}
	& D_{\alpha} (\PB_X (\XSet)  \|  \QB_X(\XSet) )\nonumber \\
	& \scalebox{1.1}{$= \frac{1}{\alpha-1} \log \sum\limits_{x \in \XSet} \frac{P_X(x)}{\sum\limits_{x \in \XSet} P_X(x)}   \Big( \frac{P_X(x)}{Q_X(x)}  \Big)^{\alpha-1}  \label{eq:D} $} \\
	& \scalebox{1.05}{$= \frac{1}{\alpha-1} \log \sum\limits_{x \in \XSet} \frac{P_X(x)}{\sum\limits_{x \in \XSet} P_X(x)}  \exp \big( (\alpha-1) D_{\alpha} (P_X(x) \| Q_X(x)) \big) $} \nonumber \\
	& \scalebox{1.1}{$= f^{-1} \Big( \sum\limits_{x \in \XSet} \frac{P_X(x)}{\sum\limits_{x \in \XSet} P_X(x)}  f \big( D_{\alpha} (P_X(x) \| Q_X(x)) \big)  \Big)$}, \nonumber
\end{align}
where $P_X(x)/\sum_{x \in \XSet} P_X(x), \forall x \in \XSet$ is a normalized probability for each $\XSet$.
The elementary information gain at each $x \in \X$ is still obtained by \eqref{eq:D} as
\begin{equation} \label{eq:DElement}
	D_{\alpha} (P_X(x) \| Q_X(x)) = \log \frac{P_X(x)}{Q_X(x)}.
\end{equation}
Note, elementary information gain is independent of $\alpha$.
It is the logarithm of Radon–Nikodym derivative \cite[eq.(6)]{Verdu2021_ErrExp_ENTROPY}
and called information lift in \cite{Ding2021_AlphaLift,Zarrabian2022_ICASSP,Hsu2019_Watchdog,Zarrabian2022_ITW,Zarrabian2023_Entropy}.
The well known R\'{e}nyi divergence expression is the definition~\eqref{eq:D} for $\XSet = \X$:
\begin{align}
	D_\alpha(\PB_X \| \QB_X)
	& = \frac{1}{\alpha-1} \log \sum_{x \in \X}  P_X(x)   \Big( \frac{P_X(x)}{Q_X(x)}  \Big)^{\alpha-1} \label{eq:Div}\\
	& = f^{-1} (\E_{P_X} [ f \big( D_{\alpha} (P_X(x) \| Q_X(x)) \big)]). \nonumber
\end{align}
This relative information quantifies the expected uncertainty in $\PB_X$ in addition to $\QB_X$, where the expectation is taken w.r.t. $\PB_X$ denoting the probability of each outcome $X = x$. Therefore, $D_\alpha(\PB_X \| \QB_X)$ is specifically called $\alpha$-order information gain in \cite{Sibson1969_InfRadius}.

\section{$\tilde{f}$-mean Information Gain}

The role of $\PB_X$ in R{\'e}nyi divergence~\eqref{eq:Div} is two-fold: the probability to be measured, where the multiplicative information gain or the exponential of elementary information gain  is raised to order $\alpha-1$: $\exp((\alpha-1)D_{\alpha} (P_X(x) \| Q_X(x))) = (P_X(x)/Q_X(x))^{\alpha-1}$; the probability that indicates the appearance frequency of each elementary information gain.

Let $\tilde{f} (t) = \exp(\frac{\alpha-1}{\alpha}t)$. We propose a new information gain measure as a $\tilde{f}$-mean, where the probability to be measured is different from frequency probability.
Assume that we want to quantify the information increase in $\boldsymbol{\Phi}_X$ from a reference probability $\QB_X$, where another probability $\PB_X$ governs how often the relative information appears at each elementary event $x$.
For each $\XSet \subseteq \X$, the $\tilde{f}$-mean information gain is
\begin{align}
	& \tilde{D}_{\alpha} \big( \boldsymbol{\Phi}_X (\XSet)  \|  \QB_X(\XSet) | \PB_X (\XSet) \big)  \nonumber  \\
	& \scalebox{1.2}{$= \frac{\alpha}{\alpha-1} \log \sum\limits_{x \in \XSet} \frac{P_X(x)}{\sum\limits_{x \in \XSet} P_X(x)}   \Big( \frac{\Phi_X(x)}{Q_X(x)}  \Big)^{\frac{\alpha-1}{\alpha}} $} \label{eq:DNew}  \\
	& \scalebox{1.1}{$= \frac{\alpha}{\alpha-1} \log \sum\limits_{x \in \XSet} \frac{P_X(x)}{\sum\limits_{x \in \XSet} P_X(x)}  \exp \big( \frac{\alpha-1}{\alpha} \tilde{D}_{\alpha} (\Phi_X(x) \| Q_X(x)) \big) $} \nonumber \\
	& \scalebox{1.2}{$= \tilde{f}^{-1} \Big( \sum\limits_{x \in \XSet} \frac{P_X(x)}{\sum\limits_{x \in \XSet} P_X(x)}  \tilde{f} \big( \tilde{D}_{\alpha} (\Phi_X(x) \| Q_X(x) ) \big)  \Big) $}, \nonumber
\end{align}
where the elementary information gain $\tilde{D}_{\alpha} (\Phi_X(x) \| Q_X(x)) = \log \frac{\Phi_X(x)}{Q_X(x)}$ equals~\eqref{eq:DElement}.\footnote{The elementary measure is always independent of $\alpha$. This is because for deterministic $Z$, the $f$-mean $\bar{Z} = f^{-1}(\E[f(Z)]) = Z$ is independent of $f$. We keep the subscript $\alpha$ in elementary measures $D_{\alpha} (P_X(x) \| Q_X(x))$ and $\tilde{D}_{\alpha} (\Phi_X(x) \| Q_X(x))$ only to show that they can be obtained by the prototype definitions \eqref{eq:D} and \eqref{eq:DNew}, respectively.}
For $\XSet = \X$, we have
\begin{align*}
 & \tilde{D}_{\alpha} \big( \boldsymbol{\Phi}_X  \| \QB_X | \PB_X \big)= \\
 & \begin{cases}
        \frac{\alpha}{\alpha-1} \log \sum_{x \in \X} P_X(x)   \Big( \frac{\Phi_X(x)}{Q_X(x)}  \Big)^{\frac{\alpha-1}{\alpha}} & \alpha \in (0,1) \cup (1,\infty),\\
        \log \min_{x\in\supp(P_X)} \frac{\Phi_{X}(x)}{Q_X(x)} & \alpha = 0, \\
        \sum_{x\in\supp(P_X)} P_X(x) \log \frac{\Phi_{X}(x)}{Q_X(x)} & \alpha = 1, \\
        \log \sum_{x \in \X} P_X(x) \frac{\Phi_{X}(x)}{Q_X(x)} & \alpha = \infty.
     \end{cases}
\end{align*}
We show in Proposition~\ref{prop:Fund} below that for given reference probability $\QB_X$ and frequency probability $\PB_X$, the maximum $\tilde{f}$-mean information gain is attained at R{\'e}nyi divergence. This proposition will be used to derive the main result Theorem~\ref{theo:Leak} in Section~\ref{sec:Leak}.
\begin{proposition}  \label{prop:Fund}
	For all $\alpha \in [0,\infty)$,
	\begin{equation}
		D_\alpha(\PB_X \| \QB_X) = \max_{ \boldsymbol{\Phi}_X }  \tilde{D}_{\alpha} \big( \boldsymbol{\Phi}_X \|  \QB_X  | \PB_X  \big)
	\end{equation}
	with the maximizer $\Phi_X^*(x) = \frac{P_{X}^{\alpha}(x)/Q_{X}^{\alpha-1}(x) }{\sum_{x} P_{X}^{\alpha}(x)/Q_{X}^{\alpha-1}(x) }$ for all $x$.
\end{proposition}
\begin{IEEEproof}
	For $\alpha \in (1,\infty)$, $\frac{\alpha-1}{\alpha} \in (0,1)$; for $\alpha \in (0,1)$, $\frac{\alpha-1}{\alpha} \in (-\infty,0)$. Then,
		\begin{align*}
		& \max_{\boldsymbol{\Phi}_X}  \tilde{D}_{\alpha} \big( \boldsymbol{\Phi}_X \|  \QB_X  | \PB_X  \big)  \nonumber  \\
		&= \max_{ \boldsymbol{\Phi}_X} \log \Big( \sum_{x \in \X} P_X(x)  \big( \frac{\Phi_X(x)}{Q_X(x)} \big)^{\frac{\alpha-1}{\alpha}}\Big)^{\frac{\alpha}{\alpha-1} }  \\
		&= \begin{cases}
			\log \big(  \max\limits_{ \boldsymbol{\Phi}_X} \sum\limits_{x\in\X} P_X(x)  \big( \frac{\Phi_X(x)}{Q_X(x)} \big)^{\frac{\alpha-1}{\alpha}} \big)^{\frac{\alpha}{\alpha-1} } & \alpha \in (1,\infty),\\
			\log \big(  \min\limits_{ \boldsymbol{\Phi}_X} \sum\limits_{x\in\X} P_X(x)  \big( \frac{\Phi_X(x)}{Q_X(x)} \big)^{-\frac{1-\alpha}{\alpha}} \big)^{\frac{\alpha}{\alpha-1} } & \alpha \in (0,1).
		\end{cases}		
	\end{align*}
	%where $(\alpha-1)/\alpha  \in (0,1)$ for $\alpha \in (1,\infty)$ and $(\alpha-1)/\alpha \in (-\infty,0)$ for $\alpha \in (0,1)$.
    In both cases, the optimization is convex programming with the solution being $\boldsymbol{\Phi}_X^*$. At extended orders,
	\begin{align*}
    	\max_{\boldsymbol{\Phi}_X} \tilde{D}_{0} \big( \boldsymbol{\Phi}_X  \| \QB_X | \PB_X \big) &= \max_{\boldsymbol{\Phi}_X} \log \min_{x\in\supp(P_X)} \frac{\Phi_{X}(x)}{Q_X(x)} \\
        & = D_0(\PB_X \| \QB_X),  \\
    	\max_{\boldsymbol{\Phi}_X} \tilde{D}_{1} \big( \boldsymbol{\Phi}_X  \| \QB_X | \PB_X \big) &=  \max_{\boldsymbol{\Phi}_X} \sum_{x\in\supp(P_X)} P_X(x) \log \frac{\Phi_{X}(x)}{Q_X(x)} \\
    	&  = D_1(\PB_X \| \QB_X), \\
    	\max_{\boldsymbol{\Phi}_X} \tilde{D}_{\infty} \big( \boldsymbol{\Phi}_X  \| \QB_X | \PB_X \big) &= \max_{\boldsymbol{\Phi}_X} \log \sum_{x \in \X} P_X(x) \frac{\Phi_{X}(x)}{Q_X(x)}  \\
    	& = \max_{x \in \X} \log \frac{P_X(x)}{Q_X(x)} = D_\infty(\PB_X \| \QB_X),
	\end{align*}
	with the maximizers $\boldsymbol{\Phi}_X^*  = \QB_X$ for $\alpha = 0$, $\boldsymbol{\Phi}_X^*  = \PB_X$ for $\alpha = 1$ and $\Phi_X^*(x)  = 1/|\argmax_x \frac{P_X(x)}{Q_X(x)}|$ if $x \in \argmax_x \frac{P_X(x)}{Q_X(x)}$ and 0 otherwise for $\alpha \rightarrow \infty$.
\end{IEEEproof}

\section{$\alpha$-Leakage: Maximum Information Gain}
\label{sec:Leak}

Information leakage is defined as an estimation problem as follows~\cite{Calmon2012_Allerton,Issa2020_MaxL_JOURNAL,Liao2019_AlphaLeak,Saeidian2023_PML}. Given a privacy-preserving channel $\PB_{Y|X}$, an input $X$ will induce a channel output $Y$ that is accessible to all users, including malicious ones. An adversary can obtain an estimation of $X$, denoted by $\hat{X}$, by applying a soft decision $P_{\hat{X}|Y}$ to $Y$.
This induces a Markov chain $X - Y - \hat{X}$.
For $\PB_X$ being the adversary's prior belief, information gain is measurable for each decision or posterior belief $\PB_{\hat{X}|Y}$.
The adversary will seek the optimal decision $\PB_{\hat{X}|Y}^*$ that maximizes the information gain, where the maximum indicates the worst-case amount of information on $X$ that is leaked to the adversary and is defined as information leakage.

For each $\PB_{X}$ and $\PB_{Y|X}$, the Sibson mutual information~\cite{Sibson1969_InfRadius}
\begin{equation*}
	I_{\alpha}^{\text{S}} (\PB_X,\PB_{Y|X}) = \frac{\alpha}{\alpha-1} \log \sum_{y \in \Y} \Big(  \sum_{x\in\X} P_{X}(x) P_{Y|X}^\alpha(y|x) \Big)^{\frac{1}{\alpha}}
\end{equation*}
is the information radius of $f$-mean R\'{e}nyi divergence.\footnote{Information radius, as defined in \cite[Sec.~2]{Sibson1969_InfRadius}, is a probability distribution that minimizes $f$-mean R\'{e}nyi divergence from a given set of probabilities. %See Appendix~\ref{app:SibsonID}.
}
The following theorem shows that the R\'{e}nyi divergence is the maximum $\tilde{f}$-mean information gain incurred at each channel output $y \in \Y$. We call it $Y$-elementary information leakage. Sibson mutual information is then interpreted as the $\tilde{f}$-mean of this $Y$-elementary information leakage.

\begin{theorem} \label{theo:Leak}
	 For all $\alpha \in [0,\infty)$,
	 	\begin{align}
	 		& D_\alpha(\PB_{X|Y=y}  \|  \PB_X) \nonumber \\
	 		&\quad\quad = \max_{ \PB_{\hat{X}|Y=y} }  \tilde{D}_{\alpha} \big( \PB_{\hat{X}|Y=y} \|  \PB_X  | \PB_{X|Y=y}  \big), \  \forall y \in \Y,  \label{eq:GainDiv}\\
	 		& I_{\alpha}^{\text{S}} (\PB_X, \PB_{Y|X}) \nonumber \\
	 		&\quad\quad =  \max_{ \PB_{\hat{X}|Y} }  \tilde{D}_{\alpha} \big( \PB_{\hat{X}|Y=y} \|  \PB_X  | \PB_{Y|X} \otimes \PB_{X}  \big), \label{eq:GainSibson}
	 	\end{align}
	 	with the maximizer
	 	$$P_{\hat{X}|Y}^*(x|y) = \frac{P_{X|Y}^{\alpha}(x|y)/P_{X}^{\alpha-1}(x) }{\sum_{x} P_{X|Y}^{\alpha}(x|y)/P_{X}^{\alpha-1}(x) }$$ for all $ (x,y) \in \X \times \Y$. In~\eqref{eq:GainSibson}, $\PB_{Y|X} \otimes \PB_{X}(x,y) = (P_{Y|X}(y|x)P_X(x) \colon (x,y) \in \X \times \Y) $.
\end{theorem}
\begin{IEEEproof}
	Equation~\eqref{eq:GainDiv} is a direct result of Proposition~\ref{prop:Fund}. For Sibson mutual information, 	
	\begin{align}
		&I_{\alpha}^{\text{S}} (X;Y) \nonumber\\
		&\scalebox{1.15}{$= \frac{\alpha}{\alpha-1} \log \E_{P_Y} \big[ \exp \big( \frac{\alpha-1}{\alpha} D_\alpha(\PB_{X|Y=y}  \|  \PB_X) \big) \big]$} \\
		&\scalebox{1.15}{$= \frac{\alpha}{\alpha-1} \log \E_{P_Y} \big[ \exp \big( \frac{\alpha-1}{\alpha}$} \nonumber \\
        & \qquad\qquad\quad  \scalebox{1.15}{$\max\limits_{ \PB_{\hat{X}|Y=y}}  \tilde{D}_{\alpha} \big( \PB_{\hat{X}|Y=y} \|  \PB_X  | \PB_{X|Y=y}  \big)\big) \big]$ } \nonumber\\
		&\scalebox{1.15}{$=\max\limits_{ \PB_{\hat{X}|Y}}   \frac{\alpha}{\alpha-1} \log \E_{P_Y} \big[ \exp \big( \frac{\alpha-1}{\alpha} $} \nonumber \\
        & \qquad\qquad\quad  \scalebox{1.1}{$\tilde{D}_{\alpha} \big( \PB_{\hat{X}|Y=y} \|  \PB_X  | \PB_{X|Y=y}  \big)\big) \big] $ }  \label{eq:GainSibsonFMean} \\
		& \scalebox{1.05}{$= \max\limits_{ \PB_{\hat{X}|Y}}  \frac{\alpha}{\alpha-1}  \log    \sum\limits_{x,y} P_{Y|X}(y|x) P_X(x) \Big( \frac{P_{\hat{X}|Y}(x|y)}{P_X(x)}  \Big)^{\frac{\alpha-1}{\alpha}} $}
		 \label{eq:MaxErrExp}
	\end{align}
	where $\PB_Y$ is the channel output probability such that $P_Y(y) = \sum_{x \in \X} P_{Y|X}(y|x) P_X(x), \forall y \in \Y$.
    The maximand in \eqref{eq:GainSibsonFMean} is a $\tilde{f}$-mean of $Y$-elementary $D_\alpha(\PB_{X|Y=y} \| \PB_X)$; the maximand in \eqref{eq:MaxErrExp}
	is a $\tilde{f}$-mean of $XY$-elementary information gain $D_\alpha(\PB_{X|Y}(y|x)  \|  \PB_X(x))$ given the frequency probability $\PB_{Y|X} \otimes \PB_X$, which is denoted by $ \tilde{D}_{\alpha} \big( \PB_{\hat{X}|Y=y} \|  \PB_X  | \PB_{Y|X} \otimes \PB_X  \big)$.
	The maximizer of \eqref{eq:MaxErrExp} is $\PB_{\hat{X}|Y}^*$ by Proposition~\ref{prop:Fund}.
\end{IEEEproof}

\subsection{Existing $\alpha$-Leakage}
\label{sec:ExistLeak}
The information-theoretic privacy leakages in~\cite{Issa2020_MaxL_JOURNAL,Liao2019_AlphaLeak,Saeidian2023_PML} are actually defined based on Markov chain  $U-X-Y-\hat{U}$, where $U$ is a sensitive attribute of input data $X$ and the adversary want to gain information on $U$.
In this case, simply substituting $U$ to $X$ in \eqref{eq:GainDiv} and  \eqref{eq:GainSibson}, we have the information leakage measures from $U$ to $Y$. They are upper bounded by the leakages from $X$ to $Y$.
\begin{corollary} \label{coro:PostProcess}
	Assume Markov chain  $U-X-Y-\hat{U}$. For all $\alpha \in [0,\infty)$,\footnote{The supremum  $\sup_{\PB_U} $ in Section~\ref{sec:ExistLeak} is over all $U$ such that Markov chain $U-X-Y-\hat{U}$ is formed for fixed $\PB_X$ and $\PB_{Y|X}$.  }
	\begin{align}
		&\sup_{\PB_U} D_\alpha(\PB_{U|Y=y}  \|  \PB_U) = D_\alpha(\PB_{X|Y=y}  \|  \PB_X) , \  \forall y \in \Y, \label{eq:PostProcessD} \\
		&\sup_{\PB_U} I_{\alpha}^{\text{S}} (\PB_U, \PB_{Y|U}) = I_{\alpha}^{\text{S}} (\PB_X, \PB_{Y|X}) . \label{eq:PostProcessSibson}
	\end{align}
\end{corollary}
The proof is omitted as Corollary~\ref{coro:PostProcess} just recites the post-processing inequality of R\'{e}nyi divergence and Sibson mutual information~\cite{Polyanskiy2010_Allerton,Verdu2015_SibsonITA}.
Similar results can be found in~\cite[Ths.1{\&}2]{Liao2019_AlphaLeak}, \cite[Th.1]{Issa2020_MaxL_JOURNAL}, \cite[Th.1]{Saeidian2023_PML} for a different notion of $\alpha$-leakage: optimal estimation decision is obtained separately for prior and posterior belief and the difference in the resulting information gain, or uncertainty reduction, defines the leakage.
According to \cite{Ding2024_ISIT}, the $\alpha$-leakage defined in~\cite{Liao2019_AlphaLeak} is\footnote{$I_{\alpha}^{\text{S}} (\PB_{U_\alpha}, \PB_{Y|U}) $ is the Arimoto mutual information~\cite{Arimoto1977}, i.e., the authors in \cite{Liao2019_AlphaLeak} actually reveal an interpretation of Arimoto mutual information in privacy leakage. }
\begin{align*}	
	\Leak_\alpha (U & \rightarrow Y) \\% &= H_{\alpha} (\PB_U) - H_{\alpha} (\PB_{U|Y}) ,  \nonumber \\
	& =  I_{\alpha}^{\text{S}} (\PB_{U_\alpha}, \PB_{Y|U}) \\
	& =  \frac{\alpha}{\alpha-1} \log \E_{P_Y} \big[ \exp \big( \frac{\alpha-1}{\alpha} D_\alpha(\PB_{U_\alpha | Y=y}  \| \PB_{U_\alpha})  \big) \big],
\end{align*}
where $\PB_{U_\alpha}$ is a scaled probability of $\PB_U$ such that  $P_{U_\alpha} (u) = \frac{P_U^{\alpha}(u)}{\sum_{u \in } P_U^{\alpha}(u)}$ for all $u \in \U$ and $P_{U_\alpha|Y}(u|y) = \frac{P_{Y|U}(y|u) P_{U_\alpha}(u)}{P_Y(y)}$ for all $(u,y) \in \U \times \Y$.
Then,
\begin{align}
	& \sup_{\PB_{U_\alpha}}  D_\alpha(\PB_{U_\alpha | Y=y}  \| \PB_{U_\alpha})  \nonumber \\
	&\qquad\qquad\qquad   =  \sup_{\PB_{U}}  D_\alpha(\PB_{U | Y=y}  \| \PB_{U}), \quad \forall y \in \Y \label{eq:DUAlpha2DU}  \\
	& \sup_{\PB_{U_\alpha}} I_{\alpha}^{\text{S}} (\PB_{U_\alpha}, \PB_{Y|U}) = \sup_{\PB_{U}} I_{\alpha}^{\text{S}} (\PB_{U}, \PB_{Y|U}) . \label{eq:Arimoto2Sibson}
\end{align}
The first equality is because $D_\alpha(\PB_{U_\alpha | Y}  \| \PB_{U_\alpha}) = \frac{1}{\alpha-1} \log \sum_{u \in \U} P_{U_\alpha}(u) P_{Y|U}^{\alpha} (y|u)/P_Y^\alpha(y)$.
Equation~\eqref{eq:Arimoto2Sibson} is the equivalence of Arimoto and Sibson mutual information when they are maximized over channel input~\cite[Th.5]{Verdu2015_SibsonITA}\cite{Csiszar1995_SibsonMutual_TIT}. It is used to prove $\sup_{\PB_U} \Leak_\alpha (U  \rightarrow Y) = I_{\alpha}^{\text{S}} (\PB_X, \PB_{Y|X}) $ via \eqref{eq:PostProcessSibson} in~\cite[Appendix~A]{Liao2019_AlphaLeak}.
Clearly from \eqref{eq:PostProcessD} and \eqref{eq:DUAlpha2DU}, the $Y$-elementary leakage is also upper bounded as $\sup_{\PB_{U_\alpha}}  D_\alpha(\PB_{U_\alpha | Y=y}  \| \PB_{U_\alpha})  = D_\alpha(\PB_{X|Y=y}  \|  \PB_X)$. This equality for $\alpha=\infty$ is proved in \cite{Saeidian2023_PML}, where $D_\infty(\PB_{X|Y=y}  \|  \PB_X)$ is called pointwise maximal leakage.

\section{Conclusion}

We proposed a $\tilde{f}$-mean information gain that quantifies the information in a probability distribution $\boldsymbol{\Phi}_X$ in addition to a reference probability $\QB_X$ conditioned on a frequency probability $\PB_X$. We proved that the maximum of the $\tilde{f}$-mean information gain is attained at R\'{e}nyi divergence between $\QB_X$ and $\PB_X$.
This result was used to propose R\'{e}nyi divergence and Sibson mutual information as $\alpha$-leakage measures.

With the cross entropy proposed in \cite{Ding2024_ISIT}, we have a pair of $\tilde{f}$-mean information measures corresponding to the existing $f$-mean measures, R\'{e}nyi entropy and divergence. We have shown in Theorem~\ref{theo:Leak} and \cite[Th.~1]{Ding2024_ISIT} that the optimization of $\tilde{f}$-mean measures give $f$-mean measures. The interplay between $f$- and $\tilde{f}$-mean should be further explored.

\bibliographystyle{IEEEtran}
\bibliography{privacyBIB_ITW2024}

\end{document}